\documentclass[useAMS,usenatbib,usedcolumn,usegraphicx]{mn2e}
\usepackage{times}
\def\kmps{km\,s\ensuremath{^\textrm{-1}}}
\def\chis{\ensuremath{\chi^2}}
\newcommand{\spec}[2]{#1\,\textsc{#2}}

\title[A search for binarity in DAO white dwarfs]
  {A search for binarity using \textit{FUSE}\ observations of DAO white dwarfs}
\author[S.A. Good et al.]
  {S.A. Good$^1$\thanks{Email: sag15@le.ac.uk},
  M.A. Barstow$^1$, M.R. Burleigh$^1$, P.D. Dobbie$^1$ and J.B. Holberg$^2$ \\
  $^1$Department of Physics and Astronomy, University of Leicester, University Road, Leicester LE1 7RH\\
  $^2$Lunar and Planetary Laboratory, University of Arizona, Tucson, AZ 85721, USA}
\date{Released 2002 Xxxxx XX}

\pagerange{\pageref{firstpage}--\pageref{lastpage}} \pubyear{2004}

\def\LaTeX{L\kern-.36em\raise.3ex\hbox{a}\kern-.15em
    T\kern-.1667em\lower.7ex\hbox{E}\kern-.125emX}

\begin{document}

\label{firstpage}

\maketitle

\begin{abstract}
We report on a search for evidence of binarity in \textit{Far-Ultraviolet Spectroscopic Explorer}\ (\textit{FUSE}) observations of DAO white dwarfs.  Spectra recorded by \textit{FUSE}\ are built up from a number of separate exposures.  Observation of changes in the position of photospheric heavy element absorption lines between exposures, with respect to the stationary interstellar medium lines, would reveal radial velocity changes - evidence of the presence of a binary system.  This technique is successful in picking out all the white dwarfs already known to be binaries, which comprise 5 out of the sample of 16, but significant radial velocity shifts were found for only one additional star, Ton\,320.  This object is also known to have an infrared excess \citep{holb05}.  
DAOs can be separated broadly into low or normal mass objects.  Low mass white dwarfs can be formed as a result of binary evolution, but it has been suggested that the lower mass DAOs evolve as single stars from the extended horizontal branch \citep{berg94}, and we find no evidence of binarity for 8 out of the 12 white dwarfs with relatively low mass.  The existence of higher mass DAOs can also be explained if they are within binary systems, but of the four higher mass stars in the sample studied, PG\,1210+533 and LB\,2 do not exhibit significant radial velocity shifts, although there were only two exposures for the former object and the latter has an infrared excess \citep{holb05}.
\end{abstract}


\section{Introduction}

DAO white dwarfs, the prototype of which is HZ\,34 \citep{koes79,wese93}, are a
group of white dwarfs that are observed to have both hydrogen and helium lines
in their optical spectra, in contrast to the more common DAs, which exhibit
only hydrogen absorption.  Hence, the presence of detectable He must arise from
the existence of a thin overlying H envelope or there must be a mixing process,
dredging up He from the deeper layers of the stellar photosphere.  Radiative
forces appear to be too low to support sufficient quantities of helium to be
able to produce the observed lines \citep{venn88}.  A thin H envelope might
arise through float up of residual H in a He-rich DO atmosphere as objects
transfer between the helium and hydrogen cooling channels \citep{font87}. 
However, the discovery by \citet{napi93} that the He\,\textsc{ii}\ line in the
optical spectra of one DAO was better reproduced by a homogeneous rather than a
layered atmospheric model was contrary to this view.  \citet{berg94}\ performed
a similar comparison on a sample of 14 DAOs, but found only one object for
which a stratified model was preferred.  Of the remaining objects, most were
found to be comparatively hot and low gravity, implying they have low mass.
Therefore, the progenitors of some of these DAOs are unlikely to have had
sufficient mass to ascend the asymptotic giant branch (AGB), and instead they
may have evolved from extended horizonal branch stars.  It was suggested that,
in these stars, weak mass loss may be occurring, a process that might be able
to support the observed quantities of helium in the line forming regions of the
DAOs \citep{ungl98,ungl00}.  Three objects (RE\,1016-053, PG\,1413+015 and
RE\,2013+400) were in close binary systems and have M dwarf (dM) companions. 
These have `normal' temperatures and gravities, yet still have detectable
helium lines in their optical spectra.  This may be because mass is lost as the
progenitor star passes through the common envelope phase, leading to the star
being hydrogen poor, and allowing a weak process, such as mass loss, to mix
helium into the line forming region of the white dwarf.  Alternatively, these
DAOs might be accreting from the wind of their companions - RE\,0720-318,
another DAO with a dM companion, has been shown to have changing helium
abundance over time, which might be due to episodic accretion
\citep{finl97,venn99}.  In addition, \citet{dobb99}\ identified likely spatial
non-uniformities in the surface abundance of helium, which are consistent with
models of accretion.  Finally, for one star in the \citet{berg94}\ sample
(PG\,1210+533), the He\,\textsc{ii}\ line profile was not reproduced
satisfactorily by either chemically homogeneous or stratified models.  In
addition, its helium line strengths have been observed to change over a time
period of $\sim$15 years \citep{berg94}.  This is not a high temperature, low
gravity object \citep{berg94,good04}, and has no infrared excess that might
indicate a companion star, so it does not appear to fit in with any of the
other DAOs.

Therefore, DAO white dwarfs form a relatively heterogeneous group.  For some, the existence of cool, non-degenerate companions may play a significant role in the presence of photospheric He, either from prior common envelope evolution or from ongoing mass transfer.  We investigate a sample of 16 DAO white dwarfs to determine how many exhibit evidence of short term radial velocity variations and thus may be members of close binary systems, using \textit{Far-Ultraviolet Spectroscopic Explorer}\ (\textit{FUSE}) data.  \textit{FUSE}\ spectra are built up from a number of exposures that last up to 90 minutes, which need to be combined to achieve the desired signal-to-noise of the proposer.  Although the signal-to-noise of these exposures is far worse than in the final combined spectrum, strong interstellar and photospheric absorption features can still be identified.  This investigation exploits this fact to search for radial velocity shifts in photospheric lines relative to the stationary interstellar medium lines, which might indicate the presence of short period binary systems.

\section{Observations}

Far-ultraviolet (far-UV) data for all the objects were recorded by the \textit{FUSE}\
spectrographs.  Table \ref{fuseobs}\ lists the observations that were used, which were downloaded by us
from the Multimission Archive (http://archive.stsci.edu/mast.html), hosted by
the Space Telescope Science Institute.  Overviews of the mission and in-orbit performance have been published by \citet{moos00} and \citet{sahn00} respectively; below is a brief description of the instrument.

\begin{table*}
 \begin{center}
 \caption{List of \textit{FUSE} observations for the stars in the sample.}
 \label{fuseobs}
 \begin{tabular}{llccccc}
 \hline
 Object & WD number & Obs. ID & Date & Length / s & Aperture & TTAG/HIST \\
 \hline
 A\,7           & WD0500-156 & B0520901000 & 2001/10/05 & 11525 & LWRS & TTAG \\
 HS\,0505+0112  & WD0505+012 & B0530301000 & 2001/01/02 & 7303 & LWRS & TTAG \\ 
 PuWe\,1        & WD0615+556 & B0520701000 & 2001/01/11 & 6479 & LWRS & TTAG \\
                & & S6012201000 & 2002/02/15 & 8194 & LWRS & TTAG \\
 RE\,0720-318   & WD0718-316 & B0510101000 & 2001/11/13 & 17723 & LWRS & TTAG \\
 TON\,320       & WD0823+317 & B0530201000 & 2001/02/21 & 9378 & LWRS & TTAG \\
 PG\,0834+500   & WD0834+501 & B0530401000 & 2001/11/04 & 8434 & LWRS & TTAG \\
 A\,31          & & B0521001000 & 2001/04/25 & 8434 & LWRS & TTAG \\
 HS\,1136+6646  & WD1136+667 & B0530801000 & 2001/01/12 & 6217 & LWRS & TTAG \\
                & & S6010601000 & 2001/01/29 & 7879 & LWRS & TTAG \\
 Feige\,55      & WD1202+608 & P1042105000 & 1999/12/29 & 19638 & MDRS & TTAG \\
 		& & P1042101000 & 2000/02/26 & 13763 & MDRS & TTAG \\
                & & S6010101000 & 2002/01/28 & 10486 & LWRS & TTAG \\
                & & S6010102000 & 2002/03/31 & 11907 & LWRS & TTAG \\
                & & S6010103000 & 2002/04/01 & 11957 & LWRS & TTAG \\
                & & S6010104000 & 2002/04/01 & 12019 & LWRS & TTAG \\
 PG\,1210+533   & WD1210+533 & B0530601000 & 2001/01/13 & 4731 & LWRS & TTAG \\
 LB\,2          & WD1214+267 & B0530501000 & 2002/02/14 & 9197 & LWRS & TTAG \\
 HZ\,34         & WD1253+378 & B0530101000 & 2003/01/16 & 7593 & LWRS & TTAG \\
 A\,39          & & B0520301000 & 2001/07/26 & 6879 & LWRS & TTAG \\
 RE\,2013+400   & WD2013+400 & P2040401000 & 2000/11/10 & 11483 & LWRS & TTAG \\
 DeHt\,5        & WD2218+706 & A0341601000 & 2000/08/15 & 6055 & LWRS & TTAG \\
 GD\,561        & WD2342+806 & B0520401000 & 2001/09/08 & 5365 & LWRS & TTAG \\
 \hline
 \end{tabular}
 \end{center}
\end{table*}

The \textit{FUSE} instrument contains four separate co-aligned optical paths
(channels), each having a mirror, a focal plane assembly and a diffraction grating, whose dispersed images share a portion of a detector.  Light from the target enters into the apertures of all
the channels simultaneously.  To obtain optimal coverage over the full hydrogran Lyman
series (apart from $\alpha$), two of the mirrors and two of the gratings are
coated with LiF over a layer of aluminium, while the others are coated with SiC
since the reflectivity of the Al+LiF is low below $\sim$1020 \AA.  Two
microchannel plate detectors (1 and 2) are used, with each subdivided into two
segments (A and B), which are separated by a small gap.  Light from a SiC and a
LiF channel falls onto each detector, resulting in 8 individual spectra. 
Thermal changes can result in rotations of the mirrors, hence most observations are
carried out with the largest available aperture (LWRS, 30 $\times$\ 30 arcsec)
to prevent the target drifting outside the aperture.  In this aperture the
point source resolution has been found to be 20\,000 for the LiF 1 channel
(\textit{The FUSE observers guide}\ -- see http://fuse.pha.jhu.edu/).   There
are two modes for recording data -- time-tagged event lists (TTAG data) or as
spectral image histograms (HIST data), which are used where the source is
bright.  As \textit{FUSE} is in a low-Earth orbit emission lines from the
Earth's atmosphere are sometimes seen, which must be dealt with during the data
analysis. A particular problem with \textit{FUSE}\ spectra is the presence of
the `worm', which is a shadow cast by the electron repeller grid located above
the detector surface, and manifests itself by a decrease in flux by up to 50
per cent, particularly in the LiF 1B segment.  The amount of flux loss varies
according to how closely the position of the grid wires coincides with the
image, and is also affected by the position of the target in the aperture and
so cannot easily be removed by the calibration software.

As a number of improvements have been made to the calibration pipeline since the data were originally processed and archived, after they were downloaded the data were reprocessed using a locally installed version of the \textsc{calfuse}\ pipeline (version 2.0.5 or later).  It was found that the wavelength calibration of each segment did not perfectly match with the others, so only data from the LiF 1A combination of mirrors and detectors were used.  This should have the best calibration since it is used in the pointing of the telescope.

The sample consists of 16 DAO white dwarfs.  Temperatures and gravities for these stars have previously been published by \citet{good04}.  In the sample, 5 are known binaries: RE\,0720-318 \citep{venn94,bars95b}, PG\,0834+500 \citep{saff98}, HS\,1136+6646 \citep{hebe96}, RE\,2013+400 \citep{poun93,bars95a}, and the double-degenerate Feige\,55 \citep{holb95}.  These stars are included in this analysis to assess the ability of the technique to detect white dwarfs in binary systems.  In addition, Ton\,320 and LB\,2 have known infrared excesses that might indicate a companion \citep{holb05}.

\section{Methodology}

Only the LiF 1A segment is used in this experiment as it is the best
calibrated.  Even so, the wavelength scale can drift between exposures; this
can be corrected for by reference to interstellar medium (ISM) lines, the
radial velocity of which should be constant.  The lines used are listed in
Table \ref{rvismlines}, with their rest frame wavelengths from \cite{mort91}.
Several of the lines were detectable in the spectra of every object.  A
Gaussian and a second order polynomial were used to model the line profile and
continuum level respectively.  The set of parameters that best reproduce the
data were determined by a least-squares fitting routine that is part of the
\textsc{idl}\ language (\textsc{curvefit}).  In addition, a 1$\sigma$ error on
the central wavelength was returned by the fitting function.  As the true
redshift of the ISM lines is not known, and the FUSE wavelength calibration is
insufficiently accurate, the mean position of the interstellar lines was taken
to be the correct absolute calibration.  The central wavelengths of
photospheric lines were then measured in a similar way.  Table~\ref{rvpholines}
lists the lines used, although again not all were present in all the objects,
or were difficult to identify in the low signal-to-noise exposures. In
addition, in lines of sight where the \spec{Ar}{i}\ ISM line at 1066.6599\,\AA\
was observed, the \spec{Si}{iv}\ was indiscernible due to line blending. 
Wavelength corrections, calculated from the positions of the ISM lines in each
exposure, were applied to each measured central wavelength, the velocity of the
lines were calculated, and finally the mean radial velocity for each exposure
was calculated.

\begin{table}
  \begin{center}
    \caption{ISM lines used to correct for wavelength calibration
    shifts between FUSE exposures, with rest frame wavelengths as
    listed by Morton (1991).}
    \label{rvismlines}
    \begin{tabular}{lr@{.}l}
      \hline
      Species & \multicolumn{2}{c}{Lab. wavelength / \AA} \\
      \hline
      \spec{O}{i}   & ~~~~~~~~~~988 & 6549 \\
      \spec{Si}{ii} & 989 & 8731 \\
      \spec{Si}{ii} & 1020 & 6989 \\
      \spec{C}{ii}  & 1036 & 3367 \\
      \spec{C}{ii}  & 1037 & 0182 \\
      \spec{O}{i}   & 1039 & 2304 \\
      \spec{Ar}{i}  & 1048 & 2199 \\
      \spec{Fe}{ii} & 1063 & 1764 \\
      \spec{Ar}{i}  & 1066 & 6599 \\
      \hline
    \end{tabular}
  \end{center}
\end{table}

\begin{table}
  \begin{center}
    \caption{Photospheric lines used to measure radial velocities.}
    \label{rvpholines}
    \begin{tabular}{lr@{.}l}
      \hline
      Species & \multicolumn{2}{c}{Lab. wavelength / \AA} \\
      \hline
      \spec{P}{v}   & ~~~~~~~~~~997 & 5240 \\
      \spec{O}{vi}  & 1031 & 9310 \\
      \spec{O}{vi}  & 1037 & 6130 \\
      \spec{Si}{iv} & 1066 & 6140 \\
      \spec{O}{iv}  & 1067 & 7680 \\
      \spec{Fe}{vii}& 1073 & 9480 \\
      \hline
    \end{tabular}
  \end{center}
\end{table}

Once the radial velocities had been measured, a test was performed to measure the significance of any differences found.  To do this, a constant velocity line was fitted to the radial velocity measurements and the value of the $\chi^2$\ statistic recorded.  Then, this was compared to a simulation of how the data would look if the radial velocity of the object were constant, but with the measurements perturbed by random errors.  To do this, randomly generated Gaussian distributed errors were calculated, the standard deviation of which was set equal to the error in each of the real radial velocity measurements, since the accuracy of the measurements depend on the length of exposure and number of lines used, and varied between measurements.  A constant velocity line was then fit to the simulated measurements and the $\chi^2$\ measured, and this was then compared to the $\chi^2$\ of the fit to the real measurements to see if the observations could be explained by measurement error alone, without having to invoke motion in a binary.  This was done 10$^6$\ times and the number of times that the simulated $\chi^2$\ exceeded that of the real $\chi^2$\ was measured.  If this occurred in less than 99.7\%\ of times, the radial velocity differences were taken to be significant at the 3$\sigma$\ level.

\section{Results}

\begin{figure}
 \includegraphics[]{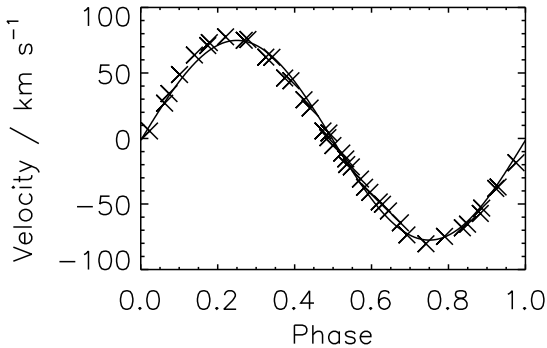}  
 \caption{\label{feige55folded} Feige 55 radial velocities folded onto the best fitting period, with a sine curve overplotted.}
\end{figure}

Table \ref{rvresults}\ lists the results of the radial velocity measurements and significance tests, and indicates whether or not the objects were already known to be in binary systems.  The technique has successfully picked out all of the known binaries, and Figure \ref{rvs_known}\ shows the radial velocity measurements for each of these stars.  The radial velocity changes are clearly evident in each, although, of these, only Feige\,55 has greater than 10 measurements, and has a complete orbit of radial velocity measurements, allowing the period and amplitude of the variations to be measured.  The best fitting sine curve to the data was determined using the \textsc{idl curvefit}\ function,  which uses a $\chi^2$\ minimisation technique to determine the best fit model parameters to a function.  The radial velocity of the white dwarf can be expressed in the following way:

$$ v(t) = v_0 + v_{max} \sin \left[ \frac{2 \pi (t - t_0)}{P} \right] $$

\noindent where the radial velocity $v$ at time $t$ is dependent on the system velocity $v_0$, the velocity semi-amplitude $v_{max}$, the system period $P$, and the epoch $t_0$.  Since the \textsc{fuse}\ observations were widely spaced in time relative to the period, the best fitting parameters, in particular the epoch, are very sensitive to the period and simultaneously fitting the period, epoch, velocity semi-amplitude and system velocity leads to unpredictable results.  Instead, the system velocity and velocity semi-amplitude were fit to the data for a range of periods and epochs, and the parameters that provided the least reduced $\chi^2$\ of all the fits were recorded.  The results are shown in Figure \ref{feige55folded}; the best fitting period, velocity semi-amplitude and system velocity are 1.493 days, 74.99 km\,s$^{-1}$\ and -0.685 km\,s$^{-1}$\ respectively.  This compares with 1.493 days, 77.4 km\,s$^{-1}$\ and 0.1 km\,s$^{-1}$\ measured by \citet{holb95}\ from optical and \textit{International Ultraviolet Explorer}\ (\textit{IUE}) data, and a period of 1.489 days, velocity semi-amplitude of 80.377 km\,s$^{-1}$\ and system velocity of -1.19 km\,s$^{-1}$\ found by \citet{kruk03}, using \textit{FUSE}\ data.  In our calculation, $t_0$, which is the reference epoch when the radial velocity equalled the system velocity and when the radial velocity is increasing, was at 2451601.762 (heliocentric Julian date).  Although the figure shows that visually the sine curve provides a good match to the data, the reduced $\chi^2$\ statistic comparing the sine curve to the data is 23, indicating that the formal errors returned by the fitting function underestimated the true errors inherent in the technique; in addition, the fact that the fit statistic does not take into account that each exposure lasts $\sim$90 minutes, rather than being discrete measurements, may be contributing to the high value.

\begin{table}
\begin{center}
    \caption{Results of the Monte Carlo analysis.  Those objects that
    the calculations suggest have real radial velocity shifts at
    the 3$\sigma$\ level are underlined.  The stars in
    bold are already known to have companions.}
    \label{rvresults}
    \begin{tabular}{lcccc}
      \hline
      Object & Number of & Best fitting & Chance of variations \\
      & exposures & vel / \kmps & occurring randomly \\
      \hline
      A\,7          & 5   & 38.8   & 0.01540 \\
      HS\,0505+0112 & 3   & 50.4   & 0.02695 \\
      PuWe\,1       & 6   & 17.5   & 0.07894 \\
      \textbf{\underline{RE\,0720-318}}  & 5   & -194.4 & 0.00000${*}$ \\
      \underline{TON\,320}      & 9   & 33.8   & 0.00000${*}$ \\
      \textbf{\underline{PG\,0834+500}}  & 3   & -46.6  & 0.00001 \\
      A\,31         & 4   & 80.8   & 0.01788 \\
      \textbf{\underline{HS\,1136+6646}} & 4   & 53.2   & 0.00000${*}$ \\
      \textbf{\underline{Feige\,55}}     & 41  & 4.2    & 0.00000${*}$ \\
      PG\,1210+533  & 2   & -3.3   & 0.93316 \\
      LB\,2         & 5   & 15.1   & 0.00882 \\
      HZ\,34        & 4   & 6.7    & 0.56639 \\
      A\,39$^{**}$     & 3   & 2.2    & 0.36889 \\
      \textbf{\underline{RE\,2013+400}}  & 9   & -15.2  & 0.00000${*}$ \\
      DeHt\,5$^{**}$   & 2   & -40.9  & 0.65238 \\
      GD\,561       & 2   & -12.5  & 0.61473 \\
      \hline
      \multicolumn{5}{l}{\textbf{*} In none of the Monte Carlo trials was the minimum \chis\ achieved by} \\
      \multicolumn{5}{l}{fitting a constant velocity line to the real data less than the \chis\ from} \\
      \multicolumn{5}{l}{fitting a constant velocity line to the simulated data.}\\
      \multicolumn{5}{l}{\textbf{**} Velocity measurements made
      difficult by H$_\mathrm{2}$\ absorption.} \\
    \end{tabular}
    \end{center}
\end{table}

\begin{figure}
 \includegraphics[]{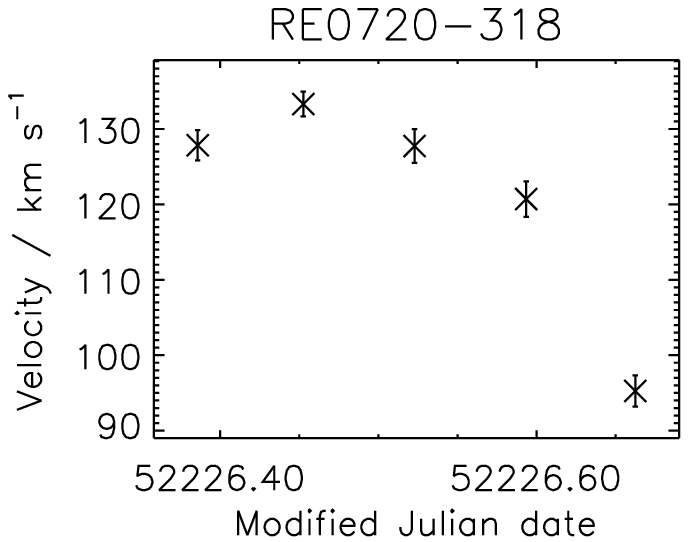} 
 \includegraphics[]{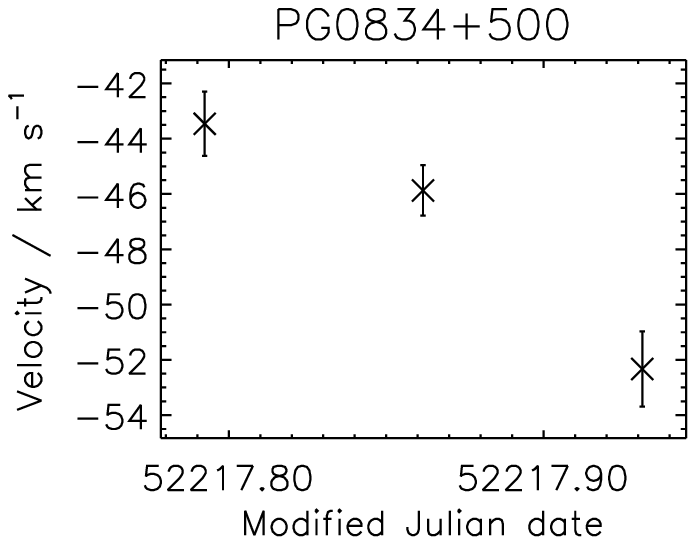} 
 \includegraphics[]{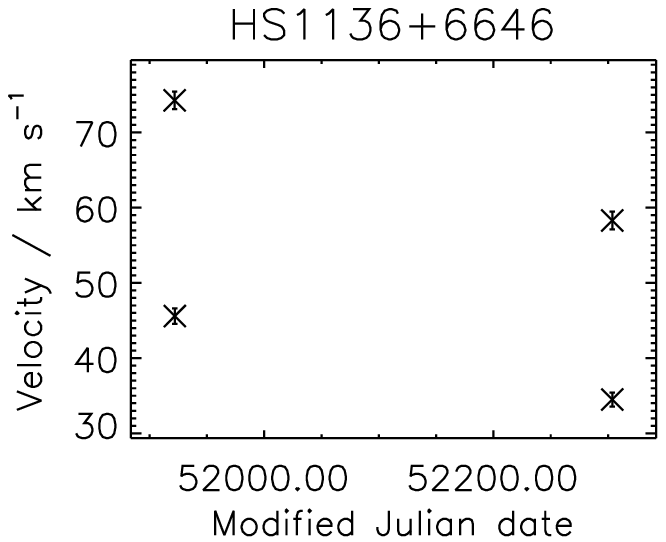} 
 \caption{\label{rvs_known} Plots of the radial velocity measurements made for the white dwarfs that are known to be in binary systems.}
\end{figure}

\begin{figure}
 \includegraphics[]{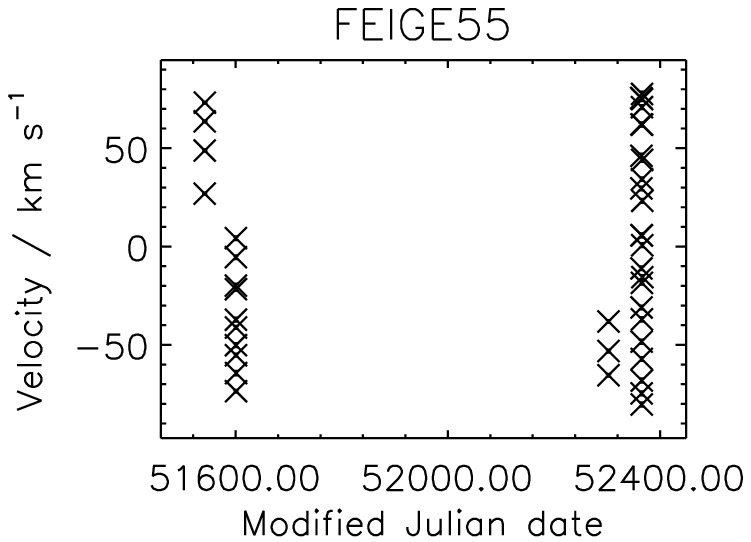} 
 \includegraphics[]{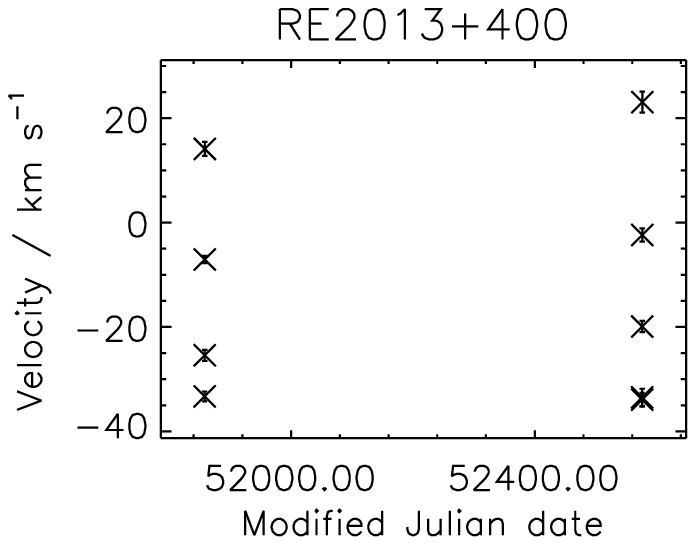} 
 \contcaption{}
\end{figure}

\begin{figure}
 \includegraphics[]{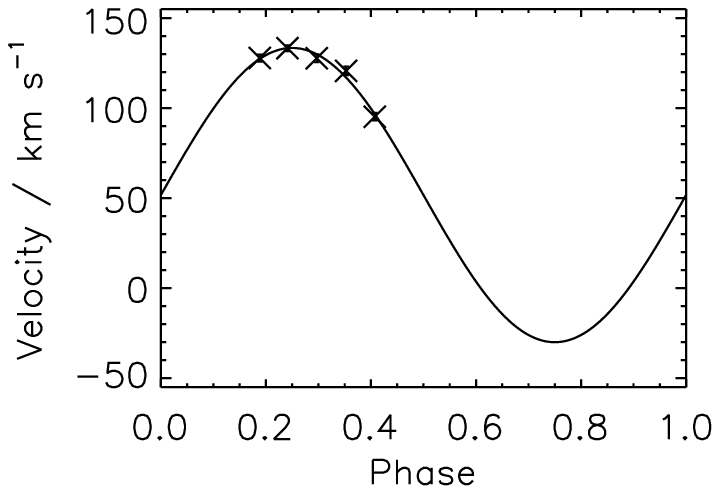}  
 \caption{\label{re0720_vennes} RE\,0720-318 radial velocities folded onto the period, with the predicted radial velocities overplotted.}
\end{figure}

Orbital parameters for RE\,0720-318 and RE\,2013+400 have been published by \citet{venn99}.  These parameters can be used to predict the radial velocity of the white dwarfs at the time of the \textit{FUSE}\ observations in order to check for consistency with our radial velocity measurements.  For RE\,0720-318, it was noted that the predicted radial velocities were out of phase with the measurements by a small amount, and that the magnitudes of the velocities were also slightly different.  To quantify these differences the set of parameters that best matched the measured quantities were found by searching around the values given by \citet{venn99}.  Predicted radial velocities were calculated for values between $\pm$3$\sigma$ of the \citet{venn99}\ quantities for the time when the velocities increase through their mean value and the velocity semi-amplitude of the orbit.  The value of the orbital period was not varied since \citet{venn99}\ quote this number to high precision, and a change of 1$\sigma$\ in the period resulted in a change in $\chi^2$\ of only $<$0.001.  Since the \textit{FUSE}\ observations sample only a limited part of the white dwarf orbit, there is insufficient information to find both the mean radial velocity and velocity semi-amplitude.  Therefore, we also did not attempt to vary the mean radial velocity from the value given by \citet{venn99}.  For each combination of parameters, $\chi^2$\ was calculated, and the set of parameters that gave the lowest $\chi^2$\ recorded.  This resulted in the following set of parameters for the orbit of RE\,0720-318: reference Julian date when the radial velocities increase through the mean, 2452226.649 (a decrease compared to the \citet{venn99}\ ephemeris of 1.38 of their sigmas), mean white dwarf radial velocity, 51.7 \kmps, velocity semi-amplitude, 81.8 \kmps\  (1.62$\sigma$\ higher than the \citet{venn99}\ value, although this will also incorporate any error in the mean velocity), and period, 1.26243 days.  The $\chi^2$\ for these parameters is 3.95, compared to 140.97 for the values given by \citet{venn99}.  Figure \ref{re0720_vennes}\ shows the measured radial velocities for RE\,0720-318 and the velocities that were predicted using these parameters.

\begin{figure}
 \includegraphics[]{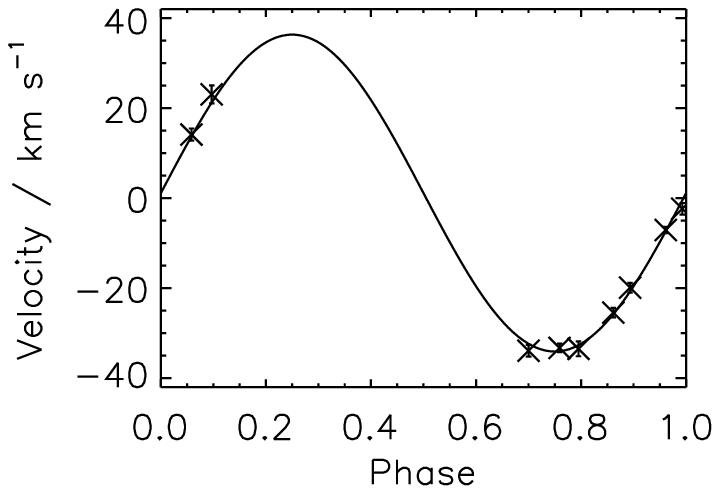}  
 \caption{\label{re2013_vennes} RE\,2013+400 radial velocities folded onto the period, with the predicted radial velocities overplotted.}
\end{figure}

The procedure described above was also followed for RE\,2013+400.  However, for this object the \textit{FUSE}\ observations cover a greater proportion of the orbit than for RE\,0720-318.  It was therefore possible to vary the period and mean radial velocity from the numbers given by \citet{venn99}\ to obtain values for these parameters also.  The set of orbital parameters found for RE\,2013+400 were: reference Julian date when velocities increase through the mean, 2451858.081 (0.35$\sigma$\ greater than the \citet{venn99}\ value), mean radial velocity, 1.1 \kmps\  (+1.26$\sigma$), velocity semi-amplitude, 35.2 \kmps\  (+2.13$\sigma$), and period, 0.705521 days (+0.29$\sigma$, although this change improved $\chi^2$\ by only 0.44).  $\chi^2$\ for this set of parameters is 6.11, compared to 35.47 for the \citet{venn99}\ values.  Figure \ref{re2013_vennes}\ shows the measured and predicted velocities for this object.  These results suggest that the radial velocities measured from \textit{FUSE}\ data are reliable and are consistent with the published orbital parameters.  Even with the limited number of radial velocity measurements available it was possible to refine the orbital parameters, suggesting that \textit{FUSE}\ data could potentially be used in the future to improve the ephemerides for stars.

\begin{figure}
 \includegraphics[]{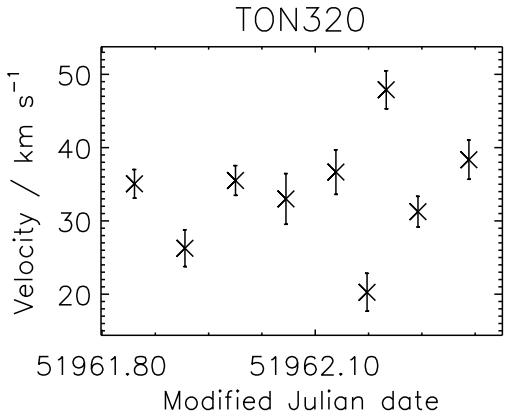} 
 \caption{\label{rvs_ton320} Plots of the radial velocity measurements made for Ton\,320, which is a suspected binary, and for which significant radial velocity changes were detected.}
\end{figure}

\begin{figure}
 \includegraphics[]{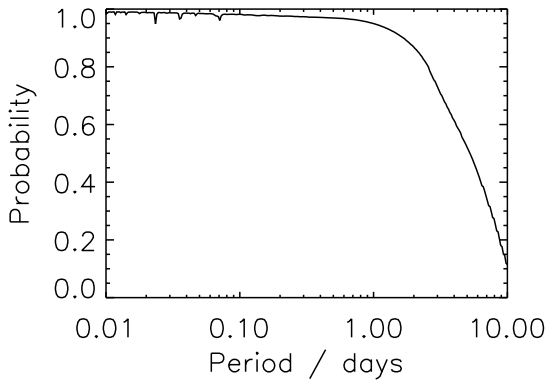} 
 \caption{\label{ton320_probs} The probability that radial velocity measurements would be detected from the data for Ton\,320.}
\end{figure}

The sixth object for which significant radial velocity changes were found was Ton\,320, which has an infrared excess and thus is suspected to be within a binary system \citep{holb05}.  The radial velocity measurements for this object are shown in Figure \ref{rvs_ton320}.  In contrast to the radial velocity measurements for the known binaries, shown in Figure \ref{rvs_known}, the radial velocities seen for Ton\,320 appear relatively constant, apart from the second, sixth and seventh measurements out of the total of nine, and no sinusoidal nature to the velocity shifts is obvious.  Therefore, it is possible that this is a spurious detection of radial velocity changes, due to an underestimation of the errors in the radial velocity measurements.  Alternatively, this object may be within a binary system, with the sinusoidal nature of the radial velocity variations hidden within the errors, or by the separation of the exposures.  To investigate this possibility, the range of periods to which the radial velocity measurement technique is sensitive for this object was first found.  To determine this, the radial velocity changes that would occur between the observations made by \textit{FUSE}\ were calculated for situations where the object was assumed to be in a binary system with a number of different inclinations and at a number of different phases at the time of first observation.  The white dwarf was assumed to have a mass of 0.5 M$_\odot$, while a conservative value of 0.1 M$_\odot$\ was assumed for the companion; a more massive companion will result in larger radial velocity shifts.  The radial velocity shift that was assumed to be detectable was set to the maximum error in the velocity measurements for the object (3.46 \kmps\ for Ton\,320).  For each period, the number of times where radial velocity changes would be detectable was counted and normalised relative to the total number of trials, with the results plotted in Figure \ref{ton320_probs}.  The plot demostrates that there is a greater than 90\%\ probability that if Ton\,320\ is within a binary system with period less than one day, radial velocity variations would be observed within the measurements.  However, this does not preclude the possibility that Ton\,320\ may be in such a binary system but with weak radial velocity variations because, for example, it is at an unfavourable inclination, making them difficult to detect.

To determine if it is possible that the radial velocity measurements do follow a sinusoidal pattern, a procedure identical to that used to determine the best fitting sine curve to the Feige\,55\ velocities was used.  The best fitting curve had the following parameters: period, 0.037 days, epoch when radial velocities increase through the mean value, 2451961.535, velocity semi-amplitude, 15.95 km\,s$^{-1}$, and system velocity, 44.99 km\,s$^{-1}$.  The reduced $\chi^2$\ of the fit is $<$4, much less than that obtained for Feige\,55, which, if the errors on the radial velocity measurements are underestimated, suggests that noise is being fitted.  In addition, this period is very short compared to, for example, the systems listed by \citet{hill00}.  However, other solutions that provide higher values of $\chi^2$\ are possible, but more closely spaced measurements are required to constrain the fit and to determine with certainty if Ton\,320 lies within a close binary system.

\begin{figure}
 \includegraphics[]{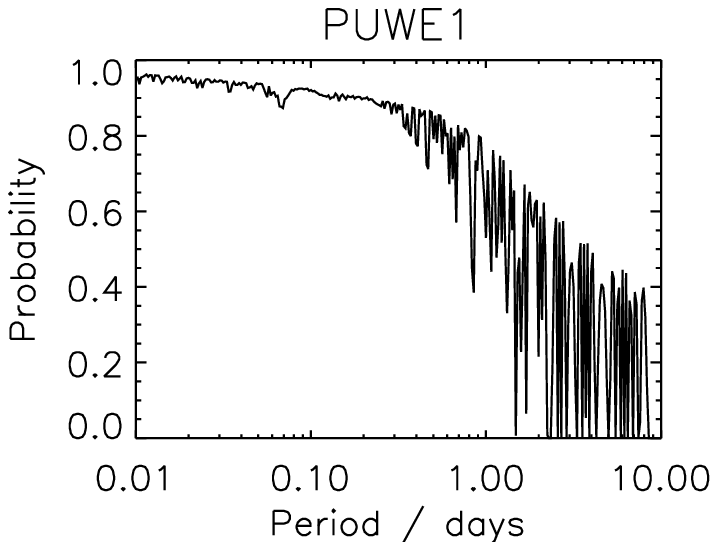}
 \includegraphics[]{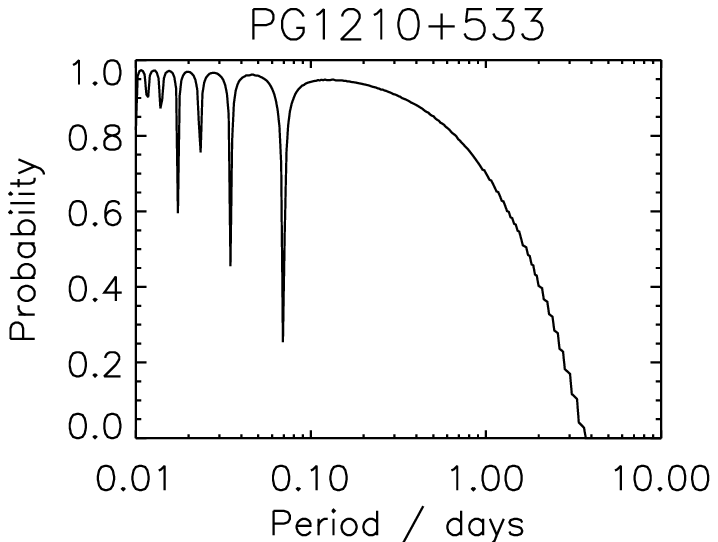}  
 \includegraphics[]{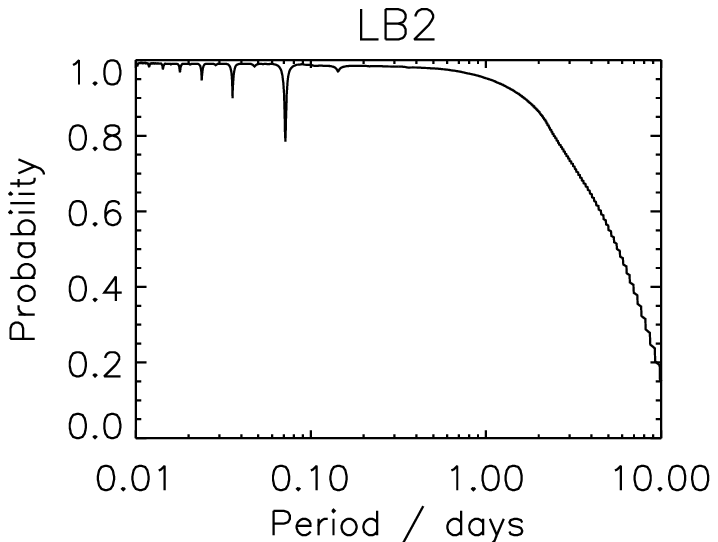}  
 \caption{\label{probs} The probability of detecting radial velocity variations for PuWe\,1, PG\,1210+533 and LB\,2.}
\end{figure}

The remaining objects are those for which no significant radial velocity changes were found.  However, the significance test for radial velocity changes was only narrowly failed by a number of these: A\,7, HS\,0505+0112, PuWe\,1, A\,31 and LB\,2 (which \citet{holb05}\ identified as having an infrared excess), for which the result of the significance test was $<$0.1.  The radial velocity measurements for these objects are shown in Figure \ref{rvs_marginal}.  Finally, radial velocity plots for PG\,1210+533, HZ\,34, A\,39, DeHt\,5 and GD\,561, for which the results of the significance tests were $>$0.1, are shown in Figure \ref{rvs_none}.  However, none of these objects have more than 6 exposures, which limits the sensitivity of the method; Figure \ref{probs}\ demonstrates the probability of detecting radial velocity shifts for periods between 0.1 and 10 days, for PuWe\,1, PG\,1210+533 and LB\,2.  PG\,1210+533 has only two exposures from a single \textit{FUSE}\ observation, and therefore our method for detecting binarity is least sensitive for this object, with the sensitivity beginning to decline above 0.1 days and strongly decreasing where the period is a multiple of the separation in time of the exposures.  For LB\,2, with five exposures, a better than 90\%\ chance of detecting radial velocity changes if it were within a binary system for period up to 1 day, and with the additional exposures, the separation in time of the measurements is less significant.  PuWe\,1 has six exposures, but these were recorded in two observations, separated in time by over a year.  The probability of detecting radial velocity changes is limited by this separation in time and, although better than for PG\,1210+533, the five continuous measurements for LB\,2 provides the best chance of detecting radial velocity shifts less than a day.  Therefore, to improve the significance of these non-detections of radial velocity shifts, longer observations are required.

\begin{figure}
 \includegraphics[]{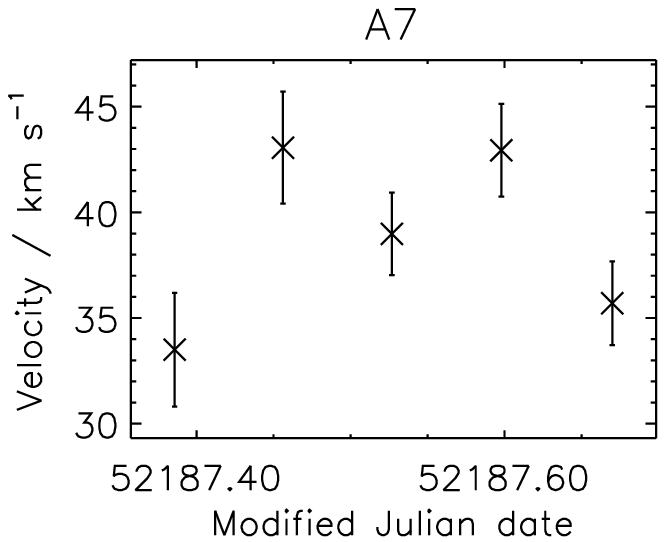} 
 \includegraphics[]{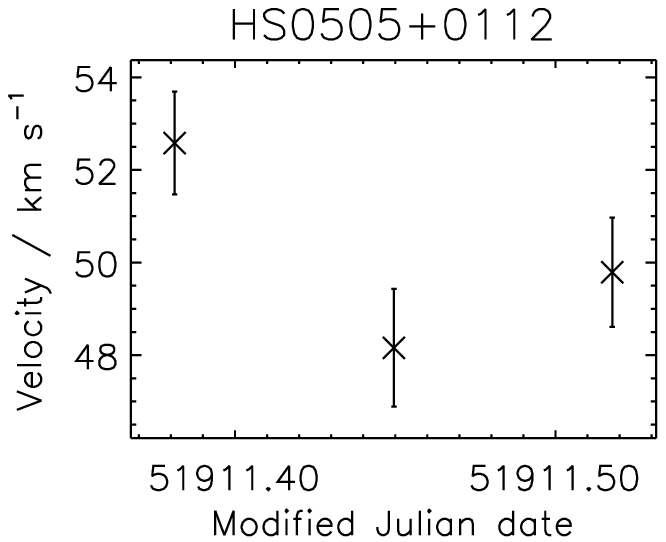} 
 \includegraphics[]{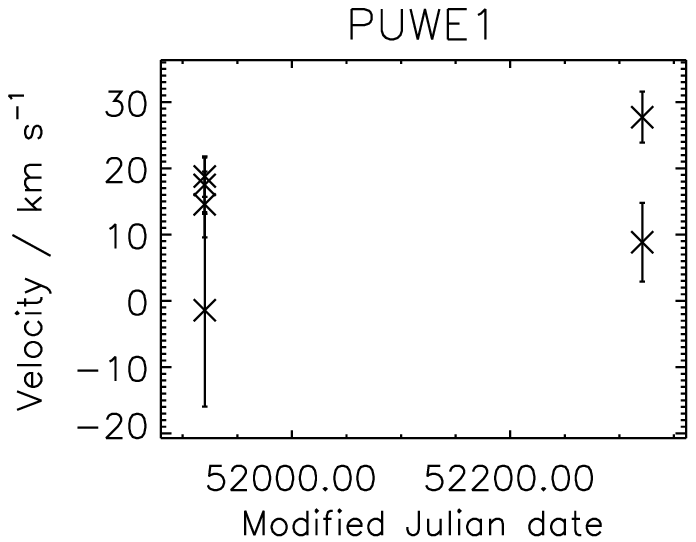} 
 \caption{\label{rvs_marginal} Plots of the radial velocity measurements made for objects where the chances of the radial velocity variations occurring randomly were less than 0.1.}
\end{figure}

\begin{figure}
 \includegraphics[]{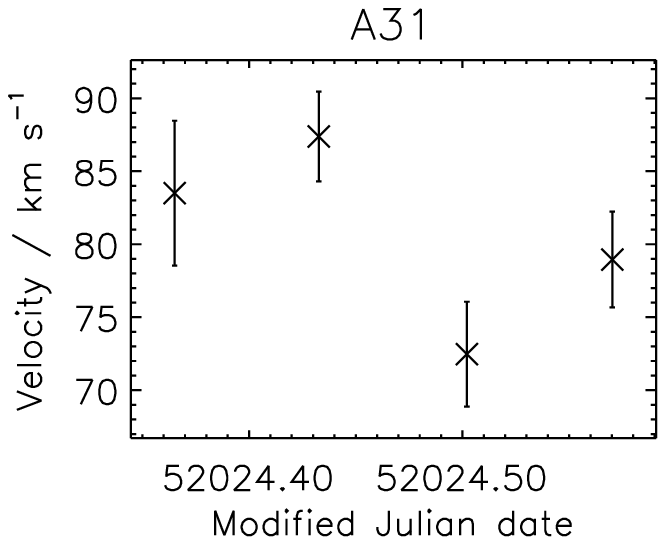} 
 \includegraphics[]{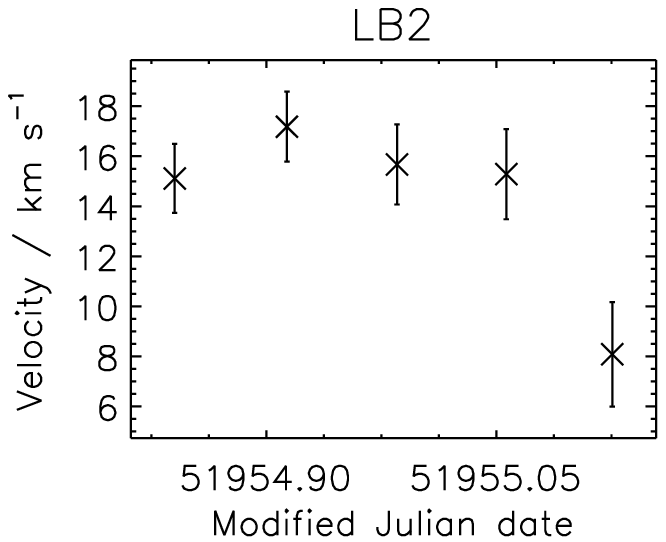} 
 \contcaption{}
\end{figure} 

\begin{figure}
 \includegraphics[]{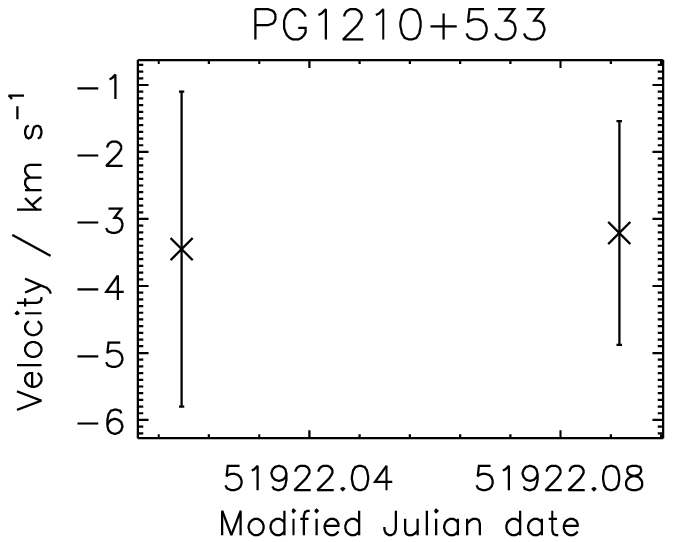} 
 \includegraphics[]{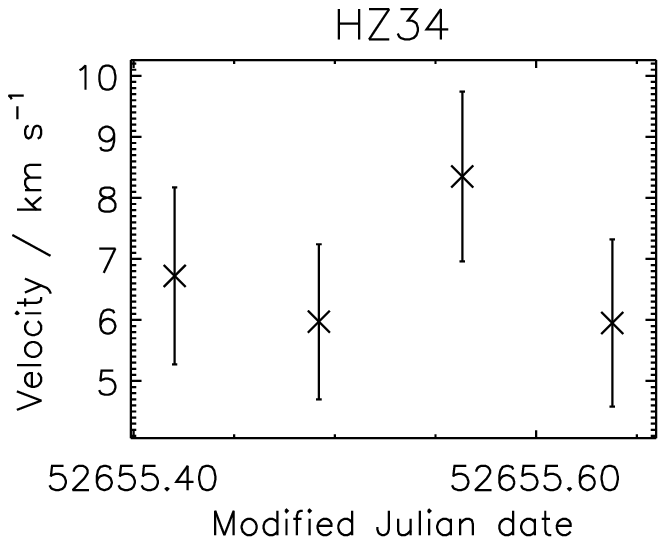} 
 \includegraphics[]{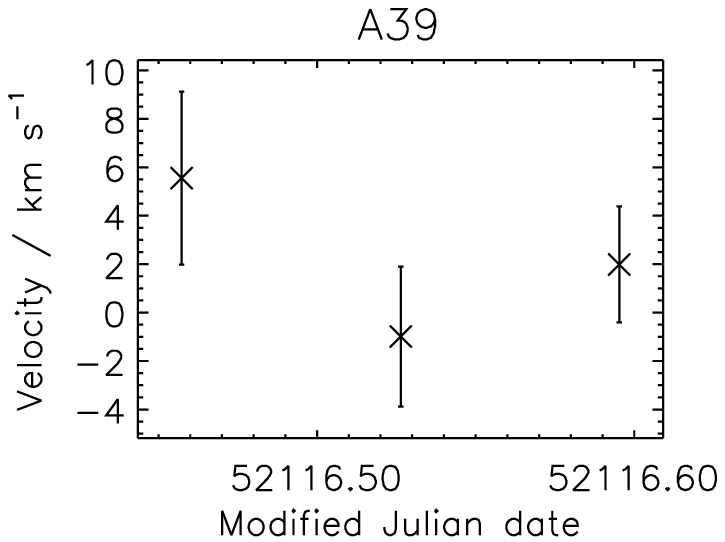} 
 \caption{\label{rvs_none} Plots of the radial velocity measurements for stars where the chances of the observed changes in radial velocity occurring randomly were greater than 0.1.}
\end{figure}

\begin{figure}
 \includegraphics[]{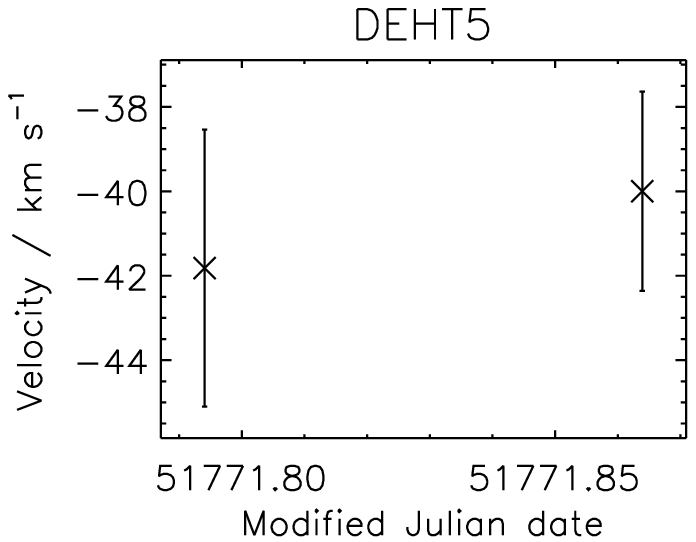} 
 \includegraphics[]{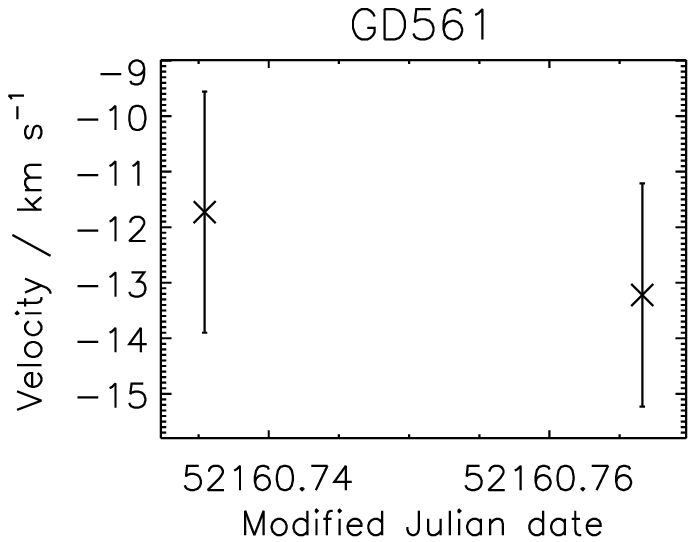} 
 \contcaption{}
\end{figure}

\section{Discussion}

Evidence for binarity has been found for only 6 DAOs.  These 6 are all known, or suspected to be in binary systems from other observations.  This suggests that the technique we have adopted is a valid way of searching for binarity in the remainder of the sample.  DAO white dwarfs can broadly be separated into two categories: `normal' mass white dwarfs, generally in binary systems, or low mass stars with possible winds.  To investigate what category each DAO falls into, we calculate the mass of each white dwarf by combining the temperatures and gravities measured by \citet{good04}\ with the evolutionary models of \citet{bloe95}\ and \citet{drie98}; these are listed in Table \ref{masses}.  Since \citet{good04}\ found differences between the best fitting model to their optical and far-UV data, we list mass estimates from both.  This introduces a large uncertainty in the mass measurements, with the \textit{FUSE}\ estimates generally higher, and in some cases up to 0.2 M$_\odot$\ different to the optical measure.  To simplify the discussion, we use the optical mass here, since the mass could not be calculated for three objects (PG\,0834+500, HS\,1136+6646 \citep[see also][]{sing04} and HS\,0505+0112) from \textit{FUSE}\ data as their temperatures were so extreme.  It should be noted that if the \textit{FUSE}\ masses are correct, most of the DAOs would have a relatively `normal' mass for a white dwarf, although they are hotter than a typical DA, and so would be post-AGB stars, rather than having to have evolved from the extended horizontal branch or in a binary system.  We divide the objects into groups; the higher mass stars all have $>$\ 0.55 M$_\odot$\ from the optical measurements, while the remainder are put into a lower mass group.

\begin{table}
\begin{center}
\caption{Mass determined by combining the temperature and gravity determinations from Balmer and Lyman line measurements of \citet{good04}\ with the evolutionary models of \citet{bloe95}\ and \citet{drie98}.}
\label{masses}
\begin{tabular}{lr@{$\pm$}lr@{$\pm$}l}
 \hline
 Object & \multicolumn{2}{c}{Balmer / M$_\odot$} & \multicolumn{2}{c}{Lyman M$_\odot$} \\
 \hline
 A\,7          & 0.508 & 0.020 & 0.673 & 0.018 \\
 HS\,0505+0112 & 0.509 & 0.017 & \multicolumn{2}{c}{-} \\
 PuWe\,1       & 0.499 & 0.027 & 0.663 & 0.060 \\
 RE\,0720-318  & 0.566 & 0.046 & 0.630 & 0.013 \\
 TON\,320      & 0.507 & 0.017 & 0.578 & 0.013 \\
 PG\,0834+500  & 0.453 & 0.020 & \multicolumn{2}{c}{-} \\
 A\,31         & 0.491 & 0.025 & 0.578 & 0.013 \\
 HS\,1136+6646 & 0.511 & 0.008 & \multicolumn{2}{c}{-} \\
 Feige\,55     & 0.439 & 0.017 & 0.518 & 0.002 \\
 PG\,1210+533  & 0.595 & 0.031 & 0.590 & 0.022 \\
 LB\,2         & 0.557 & 0.049 & 0.521 & 0.009 \\
 HZ\,34        & 0.417 & 0.019 & 0.477 & 0.032 \\
 A\,39         & 0.455 & 0.036 & 0.531 & 0.018 \\
 RE\,2013+400  & 0.642 & 0.046 & 0.661 & 0.009 \\
 DeHt\,5       & 0.470 & 0.024 & 0.412 & 0.018 \\
 GD\,561       & 0.464 & 0.029 & 0.442 & 0.023 \\
 \hline
\end{tabular}
\end{center}
\end{table}

Of the DAOs with comparitively high mass, evidence for binarity was found for RE\,0720-318 and RE\,2013+400 (both of which are DAO+dM binaries), but not for PG\,1210+533 and LB\,2.  Only two exposures were obtained for PG\,1210+533, while LB\,2 has 5, and it might be possible that radial velocity shifts were missed because the separation in time of exposures was insufficient, because the system was viewed at an unfavourable phase, or if the radial velocity measurement technique were insufficiently sensitive.  In addition, if a system is at high inclination, no radial velocity variations would be detectable.  This may indeed be the case for LB\,2, which was identified by \citet{holb05}\ as having an infrared excess.  It should also be noted that for this object the mass determined from the Lyman line results is lower than that obtained from the Balmer line results and using the lower mass would place LB\,2 in the lower mass group.  As shown in Figure \ref{probs}, since there are fewer exposures for PG\,1210+533, the method is less sensitive than for LB\,2.  However, it is still 80\%\ likely that radial velocity shifts would be detected by this experiment if it were in a binary system with period less than almost a day, unless its period happened to match the separation of exposures in time, or a multiple of that separation.  This indicates that this method is quite a sensitive way of searching for white dwarfs in binary systems, particularly if a number of exposures have been made, and hence it is unlikely, although not impossible, that PG\,1210+533 is in a short period binary system, hence the explanation for helium in the spectrum of PG\,1210+533 is still elusive.

In the observations of the remaining 12 white dwarfs, evidence of binarity was
found for only Ton\,320, PG\,0834+500, HS\,1136+6646 and Feige\,55, which are
all known or suspected binary stars.  Although low mass white dwarfs can be formed within in binary systems, these results suggest that not all of these have companions.  Therefore, for most of the high
temperature, low gravity, low mass objects, evolution within a binary system,
or accretion from a companion, cannot be used to explain their low mass or the
presence of helium absorption lines in their optical spectra, and therefore
evolution as single objects from the extended horizontal branch is still
the most likely explanation for their existence.

\section{Conclusions}

\textit{FUSE}\ data have been used to measure the radial velocity variations in 16 DAO white dwarfs, from the position of the photospheric heavy element absorption lines in their spectra.  The technique was found to be successful in detecting radial velocity variability in all the known binaries, and even with only 2 exposures there is better than a 80\%\ chance of observing velocity changes if the period of the system is less than a day, apart from where the period is a multiple of the temporal separation of exposures.

Radial velocity variations were only detected for those objects already known or suspected to be binaries, although it is possible that some of the others possess binary companions that are too highly inclined, or which were observed at unfavourable phases or have orbital periods too long with respect to the time span covered by our observations to be detected.  Of the low mass DAOs, significant changes in radial velocity were observed for only four of the twelve objects, suggesting that the majority of this type of DAO do not exist in binary systems and may therefore have evolved as single stars from the extended horizontal branch.  No evidence of binarity was found for two of the four higher mass white dwarfs in the sample.  One of these objects, LB\,2, has an infrared excess and hence may in fact have a companion.  The presence of helium in the spectrum of the other, PG\,1210+533, is still unexplained.

\section*{acknowledgments}
Based on observations made with the NASA-CNES-CSA Far Ultraviolet Spectroscopic Explorer. \textit{FUSE}\ is operated for NASA by the Johns Hopkins University under NASA contract NAS5-32985.  SAG, MAB, MRB and PDD were supported by PPARC, UK; MRB acknowledges the support of a PPARC Advanced Fellowship.  JBH wishes to acknowledge support from NASA grants NAG5-10700 and NAG5-13213.

\label{lastpage}

\end{document}